\documentclass[preprint,aps]{revtex4}
\usepackage{latexsym}
\usepackage{lineno}
\usepackage{hyperref}
\usepackage{url}
\usepackage{graphicx}
\usepackage{verbatim}
\usepackage{multirow}
\usepackage{amsmath}
\newcommand{\beq}{\begin{eqnarray}}
\newcommand{\eeq}{\end{eqnarray}}
\usepackage{mathrsfs}
\usepackage{float}
\usepackage[usenames, dvipsnames]{color}
\usepackage{mathtools}
\usepackage{slashed}
\usepackage{physics}	
\usepackage{graphicx}   
\usepackage{epstopdf}
\usepackage{subfigure}  
\usepackage{hyperref}   
\usepackage{bbold}
\usepackage{wasysym}
\usepackage{amssymb}
\usepackage{feynmp}
\usepackage[symbol]{footmisc}
\def \bs{\textbf}
\usepackage[latin1]{inputenc}
\usepackage{tikz}
\usetikzlibrary{decorations.pathmorphing}
\usetikzlibrary{shapes.misc}
\tikzset{cross/.style={cross out, draw=black, minimum size=8*(#1-\pgflinewidth), inner sep=0pt, outer sep=0pt},
cross/.default={1pt}}
\usetikzlibrary{patterns,math}
\hypersetup{
     colorlinks=true,
     linkcolor=blue,
     filecolor=blue,
     citecolor = red,      
     urlcolor=blue,
     }
     
     \definecolor{bluemathematica}{rgb}{0.368417, 0.506779, 0.709798} 
      \definecolor{orangemathematica}{rgb}{0.880722, 0.611041, 0.142051} 
     
\begin{document}

\title{Exact solution for finite center-of-mass momentum Cooper pairing}

\author{Chandan Setty$^{1,\dagger}$, Jinchao Zhao$^2$, Laura Fanfarillo$^3$, Edwin W. Huang$^2$, Peter J. Hirschfeld$^4$,  Philip W. Phillips$^2$, Kun Yang$^5$ }

\affiliation{
\\$^{1}$ Department of Physics and Astronomy and Rice Center for Quantum Materials, Rice University, Houston, Texas 77005, USA
\\$^{2}$ Department of Physics and Institute of Condensed Matter Theory, University of Illinois at Urbana-Champaign, 1110 W Green St Loomis Laboratory, Urbana, Illinois 61801, USA 
\\$^{3}$ Scuola Internazionale Superiore di Studi Avanzati (SISSA), Via Bonomea 265, 34136 Trieste, Italy
\\$^{4}$ Department of Physics, University of Florida, 2001 Museum Rd, Gainesville, Florida 32611, USA
\\$^{5}$ Department of Physics and National High Magnetic Field Laboratory, Florida State University, 1800 E Paul Dirac Dr, Tallahassee, Florida 32306, USA
\\
\\$^{\dagger}$Author to whom all correspondence should be addressed: csetty@rice.edu
}

\begin{abstract}
Pair density waves (PDWs) are superconducting states formed by ``Cooper pairs" of electrons containing a non-zero center-of-mass momentum. They are characterized by a spatially modulated order parameter and may occur in a variety of emerging quantum materials such as cuprates, transition metal dichalcogenides (TMDs) and Kagome metals. Despite extensive theoretical and numerical studies seeking PDWs in a variety of lattices and interacting settings, there is currently no generic and robust mechanism that favors a modulated solution of the superconducting order parameter in the presence of time reversal symmetry. Here, we study the problem of two electrons subject to an anisotropic ($d$-wave) attractive potential. We solve the two-body Schrodinger wave equation exactly to determine the pair binding energy as a function of the center-of-mass momentum. We find that a modulated (finite momentum) pair is favored over a homogeneous (zero momentum) solution above a critical interaction. Using this insight from the exact two-body solution, we construct a BCS-like variational many-body wave function and calculate the free energy and superconducting gap as a function of the center-of-mass momentum. A zero temperature analysis of the energy shows that the  conclusions of the two-body problem are robust in the many-body limit. Our results lay the theoretical and microscopic foundation for the existence of PDWs.  
\end{abstract}
\maketitle

\section{Introduction} 
Cooper's proof~\cite{Cooper1956} that the Fermi surface of a metal is unstable to a weak attractive potential between two quasi-particles  laid the foundation for a subsequent comprehensive theory of superconductivity in elemental metals by Bardeen, Cooper and Schrieffer~\cite{BCS1957}. The resulting bound state of the two quasi-particles, termed a ``Cooper pair", contains zero center-of-mass momentum and constitutes the basic building block of a superconductor. Recently the prospect of Cooper pairs with finite center-of-mass momentum has been raised and supported by several experimental observations in emerging quantum materials~\cite{Agterberg2019}.  In time reversal symmetric settings, such a phenomenon can manifest in pair density wave (PDW) phases --- superconducting ground states defined by a spatially varying order parameter that breaks translation symmetries of the lattice and whose real space average  vanishes. A na\"{i}ve implementation of Cooper's solution at infinitesimal coupling fails in this scenario and the following question arises: Is there an  analog of the Cooper argument for finite center-of-mass momentum pairing?  A resolution to this question would be a significant advancement toward uncovering the most basic microscopic ingredients driving Cooper pairing in PDWs.\par   
Here, we present an exact solution for the existence of such a two-body bound state with finite center-of-mass momentum in the presence of time reversal symmetry as is relevant for a PDW.
Widespread interest in PDWs has been triggered by independent experimental observations supporting short or long range PDW orders in different families of superconducting materials. These include the underdoped cuprates~\cite{Tranquada2007, Zimmerman2008, Basov2010, Basov2010-PRB, Tranquada2015, JCSDavis2016, Tranquada2018, Cavalleri2018, Hamidian2019,PJH2020, Fujita2020, Wang_arxiv2021}, Kagome metals~\cite{Guo2021} and transition metal dichalcogenides~\cite{Davis2021}. However, despite concerted theoretical~\cite{Kivelson2003-RMP, SCZhang2007, BFK2009, Tranquada2009, Paramekanti2010, Kopp2010, Barci-Fradkin2011, Kopp2011, deMelo2000, Tesanovic2011, Fradkin2012, Lee2014, Fradkin2014, Fradkin2015, Granath2017, Granath2018, Setty2021} and numerical~\cite{Ogata2002, Oles2007, Poilblanc2008, FCZhang2009, Kivelson2010, Troyer2014, PJH2020, EAKim2019, HCJiang2021, HCJiang2021-2, Yao2022, Law-Lee2019} efforts, PDWs have not been found to occur as natural ground states~\cite{HCJiang2021, HCJiang2021-2, Yao2022}. At the current time, there is no known generic and robust mechanism for why PDWs might be favored in some situations over other correlated phases including the homogeneous superconductor. Hence, a clear-cut exactly solvable model describing their origin from microscopic ingredients is absent and has presented an open problem in superconductivity for decades.  \par  

The question we address here is therefore an analog of the Cooper instability for a PDW, and our solution identifies its most basic microscopic building block from which other many-body solvable models can be constructed. While the contours of our derivation follow the original argument by Cooper~\cite{Cooper1956} -- two quasi-particles interacting via an effective attractive potential -- we instead consider a problem where the quasi-particles, with a variable center-of-mass momentum $\bs Q$, interact through an effective \textit{anisotropic} interaction. 
We then seek a solution to the Schrodinger wave equation for a two particle bound state. Our main finding is the stability of a finite center-of-mass momentum ($\bs Q \neq 0~\text{mod}~\bs G$; $\bs G$ being a reciprocal lattice vector) pair over a uniform solution with zero center of mass momentum ($\bs Q = 0$)  above a critical  interaction strength.  
Using the results of the two-body solution, we construct a variational ansatz for the ground state wave function of the many-body problem. An analysis of the superconducting pairing order parameter and free energy shows that the conclusions of the two-body solution also follow in the many-body limit. Our results thus provide a microscopic foundation for the existence of PDWs in a wide variety of superconducting materials. \par 
In the remainder of the paper, we describe the effective interaction, derive an equation for the quasi-particle bound state energy as a function of center-of-mass momentum, and finally discuss our results and conclusions.  \par
\section{Interaction} In his original work~\cite{Cooper1956}, Cooper assumed a constant (isotropic) interaction defined within an energy window set by the Debye frequency and zero otherwise. Here we instead choose an anisotropic interaction with nodes along certain directions in momentum space. We motivate the momentum structure of the interaction so that the pairing form factor reduces to the $d-$wave ($B_{1g}$) symmetry in the zero center-of-mass momentum limit as seen, for example, in the cuprates~\cite{Harlingen1995}. For the $\bs Q$ dependence of the interaction, we consider Fourier transform of the pair ``nematic" operator relevant for $d$-wave charge fluctuations as observed in Raman scattering~\cite{CuprateNematic2019}. This implies that the incoming and outgoing momentum dependence of the pair potential is factorized as a product of two scalar form factors, $f_{\bs k \bs Q}$, which are functions of the two independent variables: the relative momentum $\bs k$ and center-of-mass momentum $\bs Q$. Writing out the interaction in momentum space, we have 
\beq
V_{\bs k \bs k'} (\bs Q) = 
\begin{cases}
  -V_0~f_{\bs k \bs Q}~f_{\bs k' \bs Q}  &  \text {$E_F \leq \epsilon_{\bs k}
  \leq \Lambda$} \\
  0 & \text{otherwise},
  \end{cases}
  \label{Eq:Cases}
\eeq
with $V_0>0$, the form factor $f_{\bs k \bs Q} \equiv (h_{\bs k - \bs Q/2} + h_{\bs k + \bs Q/2}  )$, E$_F$ is the Fermi energy, $\epsilon_{\bs k}$ is the non-interacting dispersion and $\sqrt{\Lambda}$ is an ultraviolet cut-off for the relative momentum $\bs k$. For the purposes of illustration, we choose the function $h(\bs k) = \sum_{n\in \mathbb{Z}}b_n\left(\cos n k_x - \cos n k_y \right)$ as is relevant for the cuprate square lattice; $b_n$ is a constant. In the continuum, we use the appropriate expansion $h(\bs k) = \frac{1}{2 \Lambda }\sum_{n\in \mathbb{Z}}n^2 b_n (k_y^2 - k_x^2) $.  In two dimensions, a bound state is possible even without a Fermi surface in which case E$_F$ can be set to zero.  We will point it out explicitly when this is the case.  Note that the interaction is constrained only in the relative momentum via $E_F \leq \epsilon_{\bs k} 
\leq \Lambda$ while the center-of-mass momentum $\bs Q$ is treated as a variational variable with respect to which the bound state energy gain in maximized; the constraint itself is independent of the center-of-mass momentum. It is also worth mentioning that chosen momentum constraint is distinct from the interaction considered by Cooper which is defined within a strict energy window above the Fermi energy. \par 
%
\section{Results}
\subsection{The Cooper problem for a PDW:} We now use the two body interaction $V_{\bs k, \bs k'} (\bs Q)$ in the two-body Schrodinger equation. For two electrons with kinetic energy $\epsilon_{\pm \bs k + \bs Q/2}$ and total center-of-mass momentum $\bs Q$, the momentum space two-electron Schrodinger equation is written as  
\beq
(\epsilon_{\bs k + \bs Q/2} + \epsilon_{-\bs k + \bs Q/2} - E)~g(\bs k) = - \sum_{\bs k'} V_{\bs k \bs k'}(\bs Q) g(\bs k'). 
\label{Eq:SWE}
\eeq
The wave vector $\bs k$ ($\bs Q$) is the Fourier transform variable of the relative (center-of-mass) position coordinate $\bs r = \bs r_1 - \bs r_2 $ ($\bs R = \bs r_1 + \bs r_2 $) with $\bs r_i$ being the position vector of the $i$th electron. 
The variable $E$ is the eigen energy whose functional form with respect to $\bs Q$ is to be determined. $g(\bs k)$ is the Fourier component of the spatial part of the total wave function that depends on the relative position $\bs r$. To define this function more precisely, we denote the two-particle wave function as $\Psi(\bs r_1, \uparrow, \bs r_2, \downarrow )$ and decompose it into the spatial $(\psi)$ and spin components $(\eta)$ as $\Psi(\bs r_1, \uparrow, \bs r_2, \downarrow ) = \psi(\bs r_1, \bs r_2) \eta_s (\uparrow, \downarrow)$.  Here $\eta_s$ is a spin singlet between the two electrons. 
For the spatial component of the wave function, we make the ansatz for a single wave vector $\bs Q$ as $\psi(\bs r_1, \bs r_2) = \Phi(\bs r) e^{i \bs Q\cdot \bs R}$. Here $\Phi(\bs r)$ depends only on the relative coordinate $\bs r$.  Fourier decomposition of the function $\Phi(\bs r)$ defines the function $g(\bs k)$ as
$\Phi(\bs r) = \frac{1}{\sqrt v}\sum_{\bs k} g(\bs k) e^{i \bs k \cdot \bs r}$
where $v$ is the normalization volume. 
Substituting the interaction Eq.~\ref{Eq:Cases} into the momentum space Schrodinger wave equation Eq.~\ref{Eq:SWE} and using the factorization property of the interaction (See Methods), we obtain the condition
\beq
1 = |V_0| \sum_{\bs k} \left[ \frac{f_{\bs k, \bs Q}^2}{- E + \epsilon_{\bs k + \bs Q/2} + \epsilon_{-\bs k + \bs Q/2}}\right] \equiv F(E).
\label{Eq:ConditionForE}
\eeq
We can now solve Eq.~\ref{Eq:ConditionForE} for the binding energy $E(\bs Q)$. For the purposes of this discussion and relevant to the materials discussed above, we restrict our calculations to two dimensions. Generalization to three dimensions can be done readily. \par
\subsection{Continuum case} 
\begin{figure}
    \centering
    \includegraphics[width=0.7\linewidth]{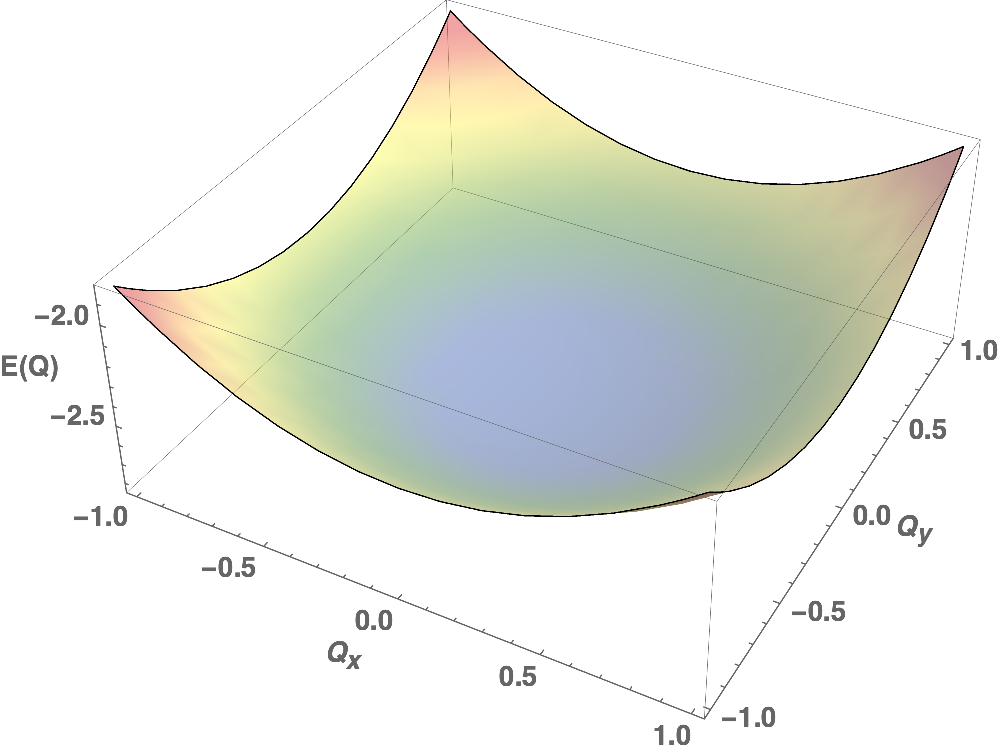} \qquad 
    \includegraphics[width=0.7\linewidth]{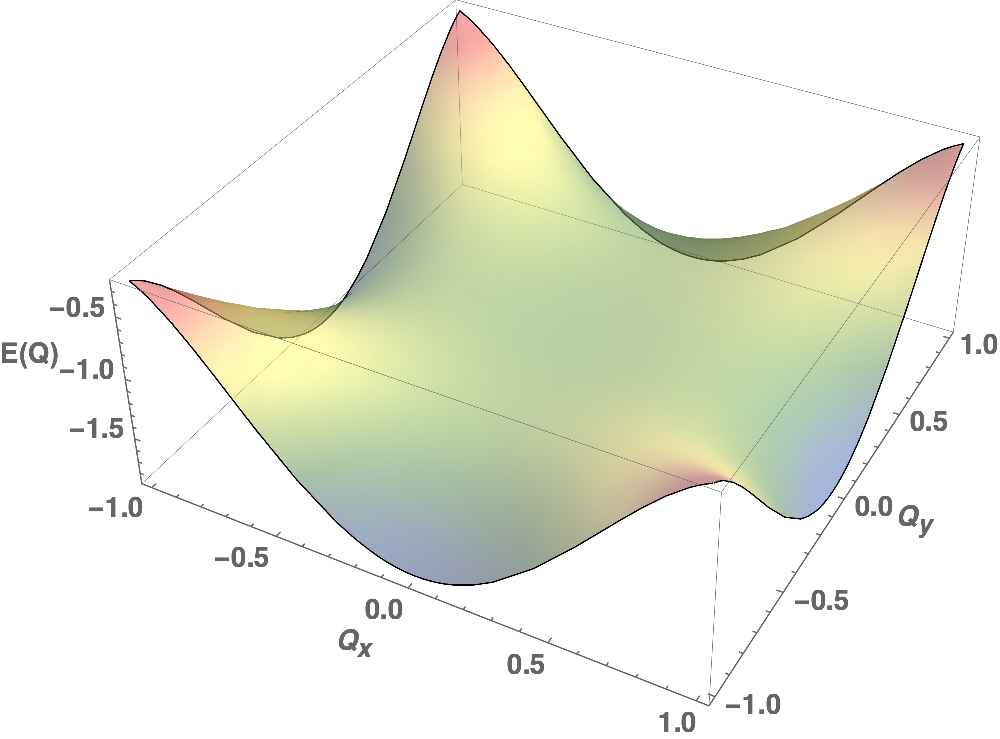} \qquad
    \caption{Binding energy $E$ as a function of momentum $\bs Q$ for isotropic $s$-wave (top) and anisotropic $d$-wave (bottom) interactions for a quadratic (continuum) dispersion. The interaction is fixed at $\nu V_0 = 4/4\pi$. The momenta are specified in units of $\sqrt{\Lambda}$ although they are unbounded in magnitude.  }
    \label{fig:Compare}
\end{figure}

We begin with the case of a continuum dispersion ($\frac{\hbar^2}{2m} =1$; $m$ is the bare electron mass) $\epsilon_{\bs k} = k^2$ where $k=|\bs k|$ and the density of states per spin is $\nu = \frac{1}{4 \pi}$. For the moment, we set the Fermi energy to zero and integrate Eq.~\ref{Eq:ConditionForE} over all allowed $\bs k$  such that only the lowest harmonic ($n=1$) in the interaction is non-zero within the window $0 \leq k^2 \leq \Lambda$. 
A plot of $E(\bs Q)$ appears in Fig.~\ref{fig:Compare} for $\nu V_0 = 4/4\pi$. For the case when the interaction is isotropic with  $f_{\bs k, \bs Q}^2 = 1$ ($s$-wave scenario), the minimum of $E(\bs Q)$ occurs at $\bs Q =0$. Hence, Cooper pairs with zero center-of-mass momentum are stabilized and any finite momentum pairing reduces the binding energy of the pair. In this scenario, the homogeneous superconductor is favorable. For the case when the interaction is anisotropic with $f_{\bs k, \bs Q}^2$ defined in Eq.~\ref{Eq:Cases}, the homogeneous solution becomes destabilized (Fig.~\ref{fig:Compare} bottom panel). While the binding energy continues to be suppressed along the diagonal (nodal) directions, the Cooper pairs acquire an unbounded gain in binding energy along the horizontal (anti-nodal) directions by taking on arbitrarily large center-of-mass momenta.  Since, in the continuum, such arbitrarily large center-of-mass momentum values are permitted, a stable minimum at finite and bounded $\bs Q$ does not exist.  Even on a lattice where the binding energy is finite and periodic, as we will see below, a non-trivial and finite center-of-mass pairing is not guaranteed. This is because the binding energy minimum can, in principle, occur for a $\bs Q$ located at a reciprocal lattice vector.   Nevertheless, such a ``run-away" binding energy is a strong indicator of non-homogeneous and modulated pairing, and understanding the stability of such solutions requires a background lattice.\par
Before we discuss the lattice case below, a few remarks are in order. Note that the interaction chosen in Eq.~\ref{Eq:Cases} has a sharp ultraviolet cut-off in momentum at $\sqrt{\Lambda}$; however, a smoothly varying interaction can also destabilize the homogeneous solution as long as it is dominant within the scale $\sqrt{\Lambda}$. Moreover, the aforementioned results in  two dimensions are robust to the inclusion of a non-zero Fermi momentum within the integration limits appearing in Eq.~\ref{Eq:ConditionForE}. Finally, while the existence of an ultraviolet cut-off aids the finite $\bs Q$ analysis, it is the non-triviality of the anisotropic interaction that really \textit{drives} the destabilization of the homogeneous pairing solution and not the nature of the chosen cut-off (see Methods for an approximate formula of the binding energy where this iss explicit). This can also be evidenced by the fact that the same choice of cut-off does not destabilize the homogeneous solution for a fully isotropic interaction. \par  
\subsection{Lattice case} 
\begin{figure}
    \centering
     \includegraphics[width=0.46\linewidth]{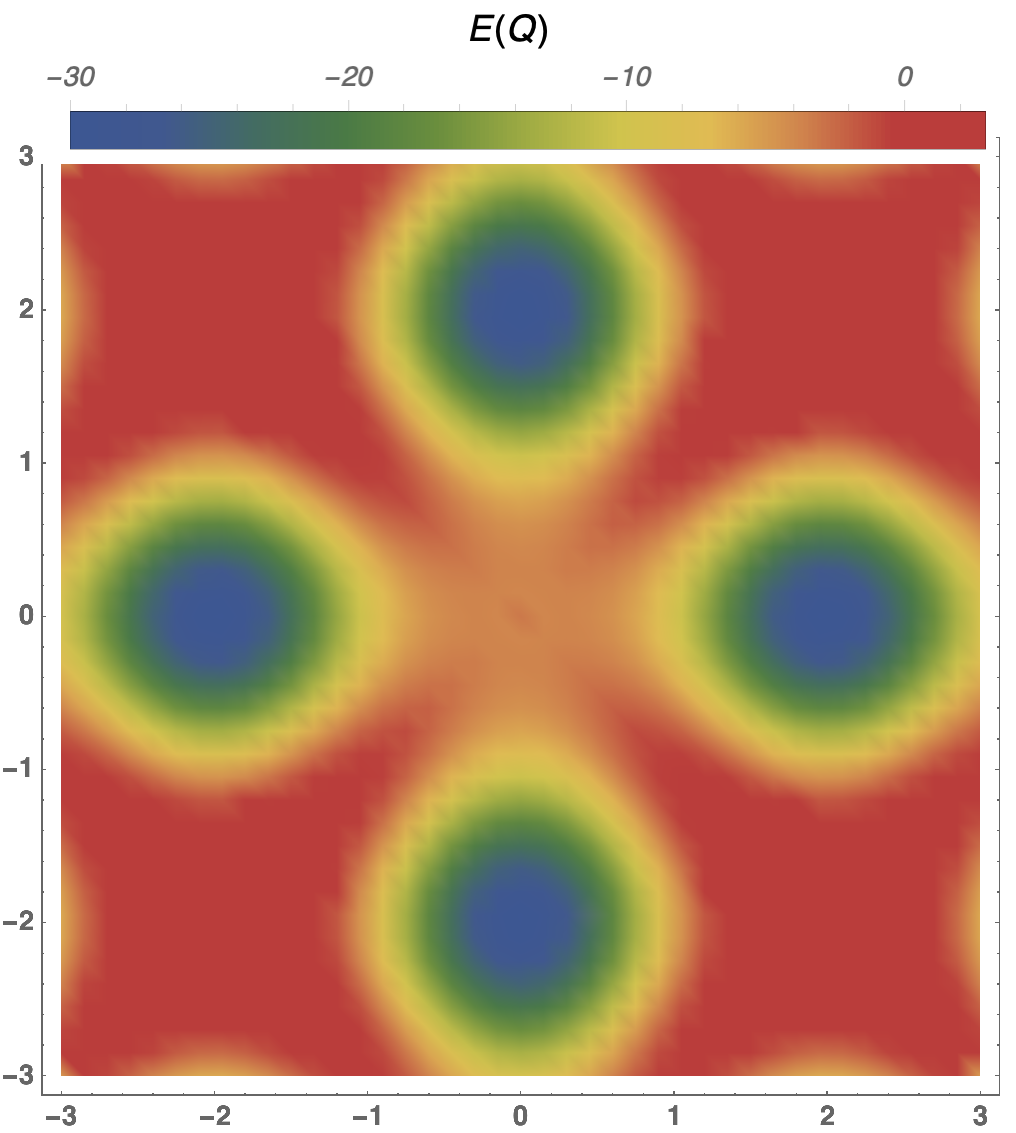} \qquad
      \includegraphics[width=0.46\linewidth]{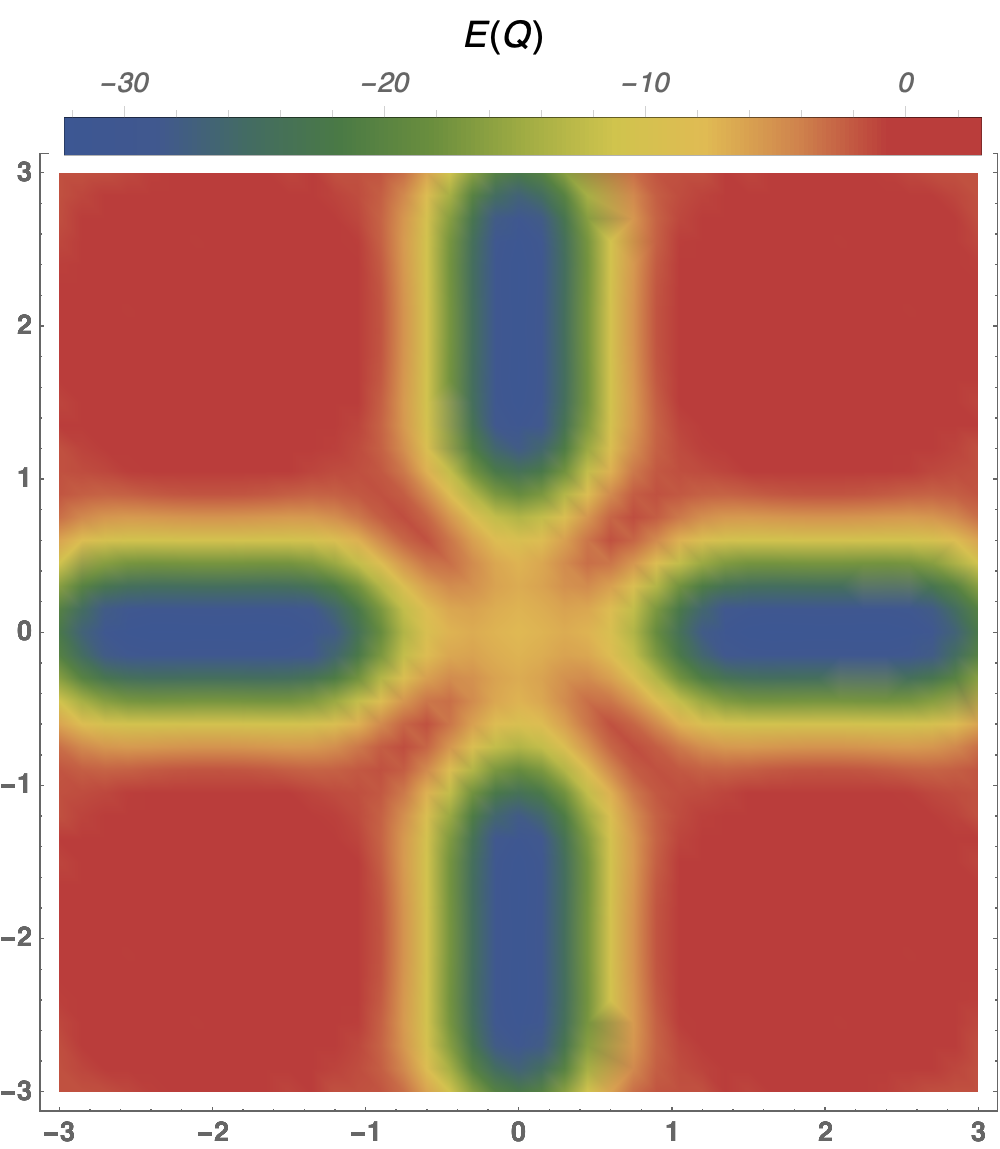} \qquad
       \includegraphics[width=0.46\linewidth]{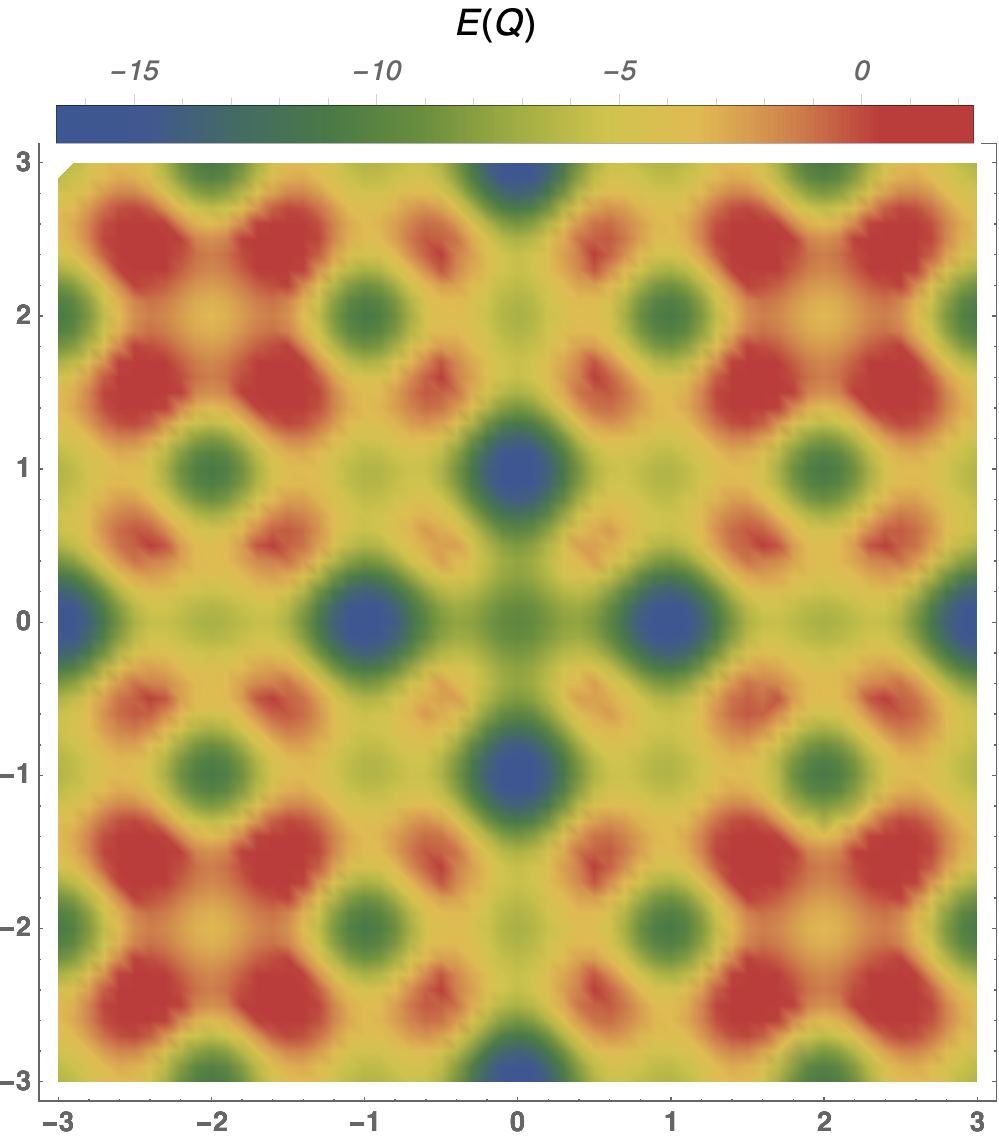} \qquad
        \includegraphics[width=0.46 \linewidth]{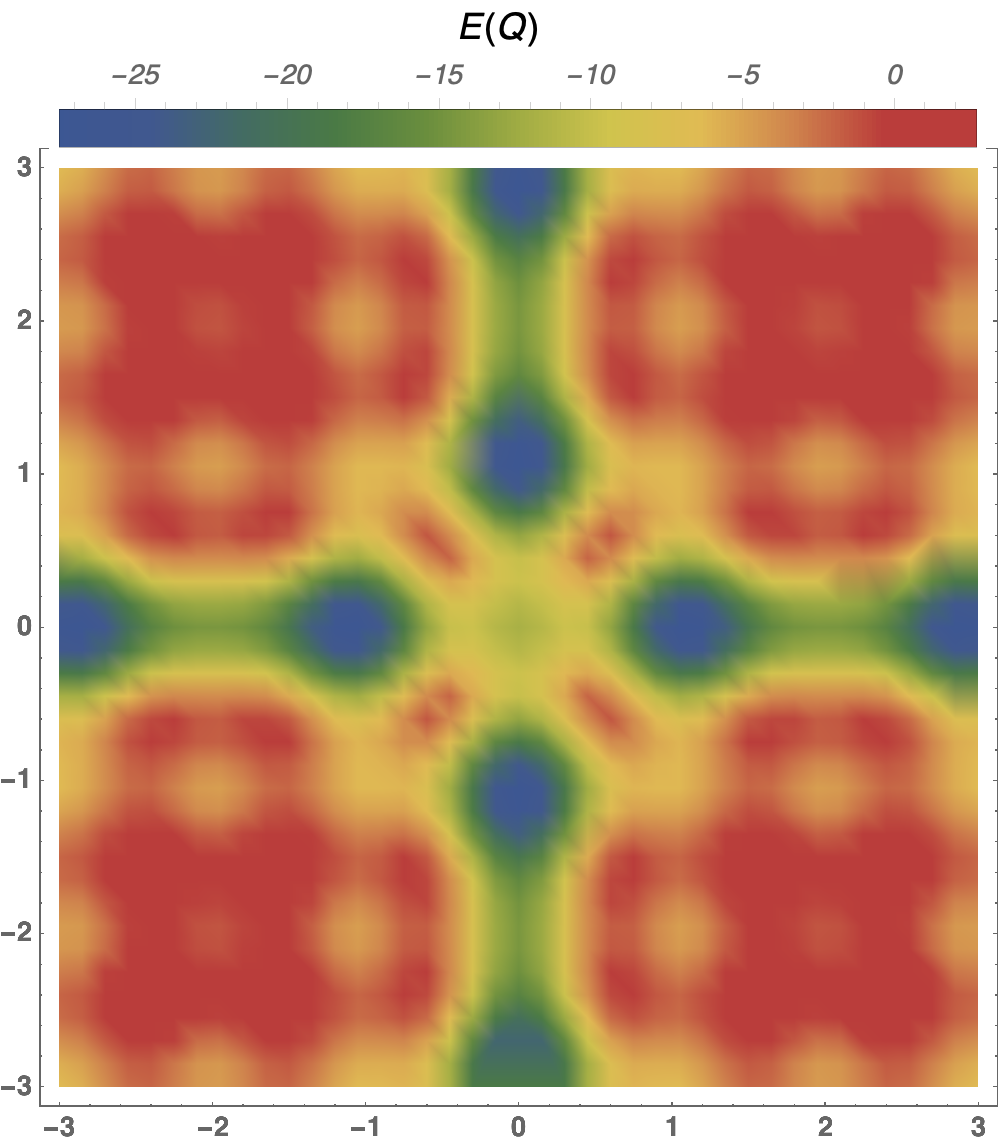} \qquad
    \caption{Color density plots of the binding energy on a lattice as a function of $\bs Q = (Q_x, Q_y)$ (in units of $\pi/a$) for various $d$-wave harmonics. (Clockwise from top left) $(b_1, b_2) = (1,0), (1,0.5), (0.5,1), (0,1)$. We have set $t=1$, $\nu V_0 = 30/2\pi$ and $\Lambda/W = 0.125$. The minimum continuously shifts from $(\pm 2\pi, 0), (0, \pm 2 \pi)$ to $(\pm \pi, 0), (0, \pm \pi)$ }
    \label{fig:HigherHarmonics}
\end{figure}

In the previous section we showed that, in continuum, the homogeneous pairing solution can become unstable to an anisotropic attractive interaction of the form appearing in Eq.~\ref{Eq:Cases}, with a ``run-away" binding energy along the anti-nodal direction. We now examine the stability of the finite $\bs Q$  phase in the presence of a lattice. We take the example of a square lattice with a dispersion $\epsilon_{\bs k} = -t (\cos k_x + \cos k_y)$ with $t=1$ and a bandwidth $W= 4 t$. The Brillouin zone extends from $[-\pi, \pi]$ in the $k_x$ and $k_y$ directions. For the function $h(\bs k)$ appearing in the interaction, we consider the two lowest $d$-wave harmonics $n=1,2$ with $\nu V_0 = 30/2\pi$ and $\Lambda/W = 0.125$. Fig.~\ref{fig:HigherHarmonics} shows a density plot of the binding energy as a function of the center-of-mass momentum $\bs Q$ for various values of $(b_1, b_2)$. The $4 \pi$ periodicity of the binding energy with respect to $\bs Q$ is set by the definition of the momentum shift in Eqs.~\ref{Eq:Cases},~\ref{Eq:SWE}, which in our case is $\pm \bs Q/2$.  With only the lowest $d$-wave harmonic $(b_1, b_2) = (1, 0)$ (top left), the minimum occurs at $\bs Q = (\pm 2 \pi, 0), (0, \pm 2 \pi)$ which are reciprocal lattice vectors of the square lattice. On the other hand, with only the second $d$-wave harmonic $(b_1, b_2) = (0, 1)$ (bottom left), the minimum occurs at $\bs Q = (\pm \pi, 0), (0, \pm  \pi)$ signalling a stable (non-trivial) finite momentum pairing instability. For intermediate values of $(b_1, b_2)$, the minima continuously shift between the two momenta (right panels).     \par
\begin{figure}
    \centering
    \includegraphics[width=0.45\linewidth]{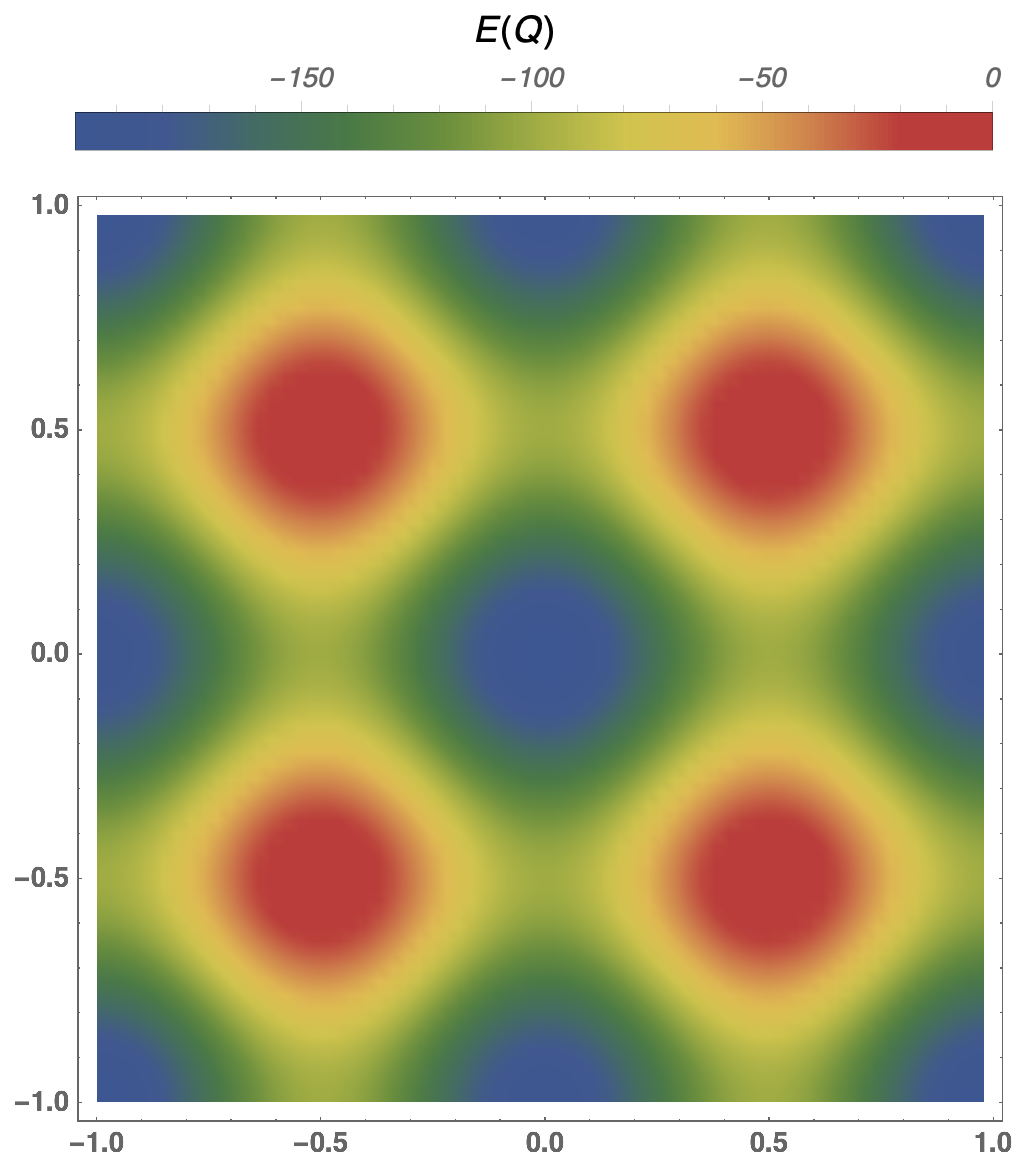} \qquad 
    \includegraphics[width=0.45\linewidth]{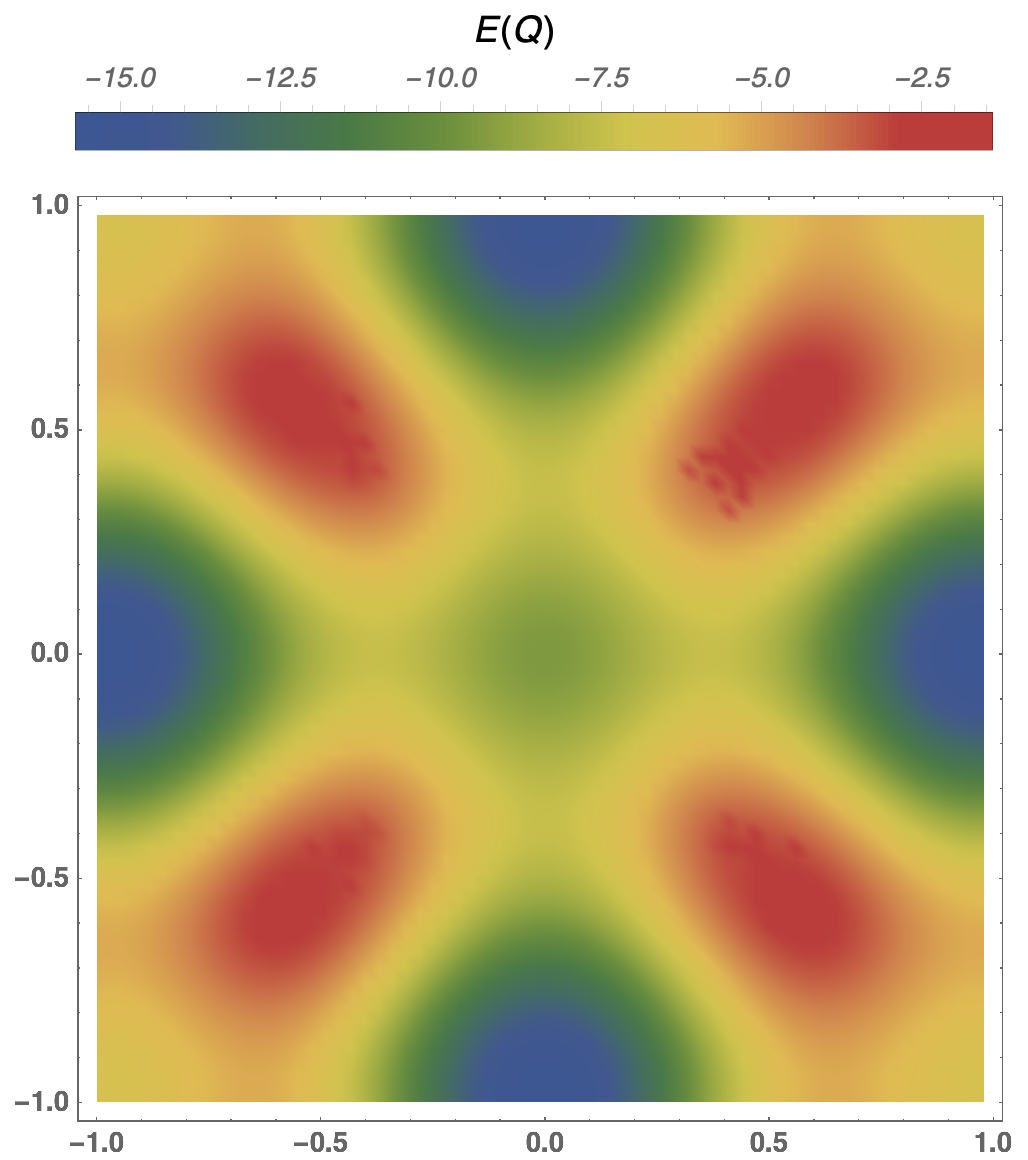} \qquad 
    \caption{Binding energy $E$ as a function of momentum $\bs Q$ (units of $\pi/a$) for second harmonic anisotropic ($d$-wave) interactions on a lattice as a function of interaction cut-offs. Left and right panels correspond to $\Lambda/W = 1.0, 0.125$ respectively. The interaction parameter is set at $\nu V_0 = 30/2\pi$ 
    }
    \label{fig:Cutoff}
\end{figure}

 
The presence of a lattice also offers an avenue to examine the role of the cut-off  $\Lambda$ on the binding energy landscape. 
Fig.~\ref{fig:Cutoff} shows a density plot of the binding energy $E(\bs Q)$ throughout the Brillouin zone for two values of the ratio $\Lambda/W$. The interaction parameter is fixed as $\nu V_0 \simeq 30/2\pi$ for the second $d$-wave harmonic and zero otherwise.  For the case when the interaction window is a large fraction of the bandwidth, the minimum of $E(\bs Q)$ occurs at $\bs Q = 0$. However, upon reducing the interaction window, there is a transition into a finite $\bs Q$ phase where the minimum occurs at the edges of the Brillouin zone at $\bs Q = (\pm \pi, 0), (0, \pm \pi)$. The transition point is non-universal and, for the chosen interaction parameter values, occurs at around $\Lambda/W \sim 0.2$ where the local minima at the Brillouin zone edges is equal to that at the center. At this value, there is a first order transition to the finite $\bs Q$ phase as a function of $\Lambda/W$ . Moreover, the cut-off chosen can be a smooth function of energy without qualitatively affecting  the transition.   \par
\subsection{Dependence on $V_0$} 
\begin{figure}
    \centering
     \includegraphics[width=1\linewidth]{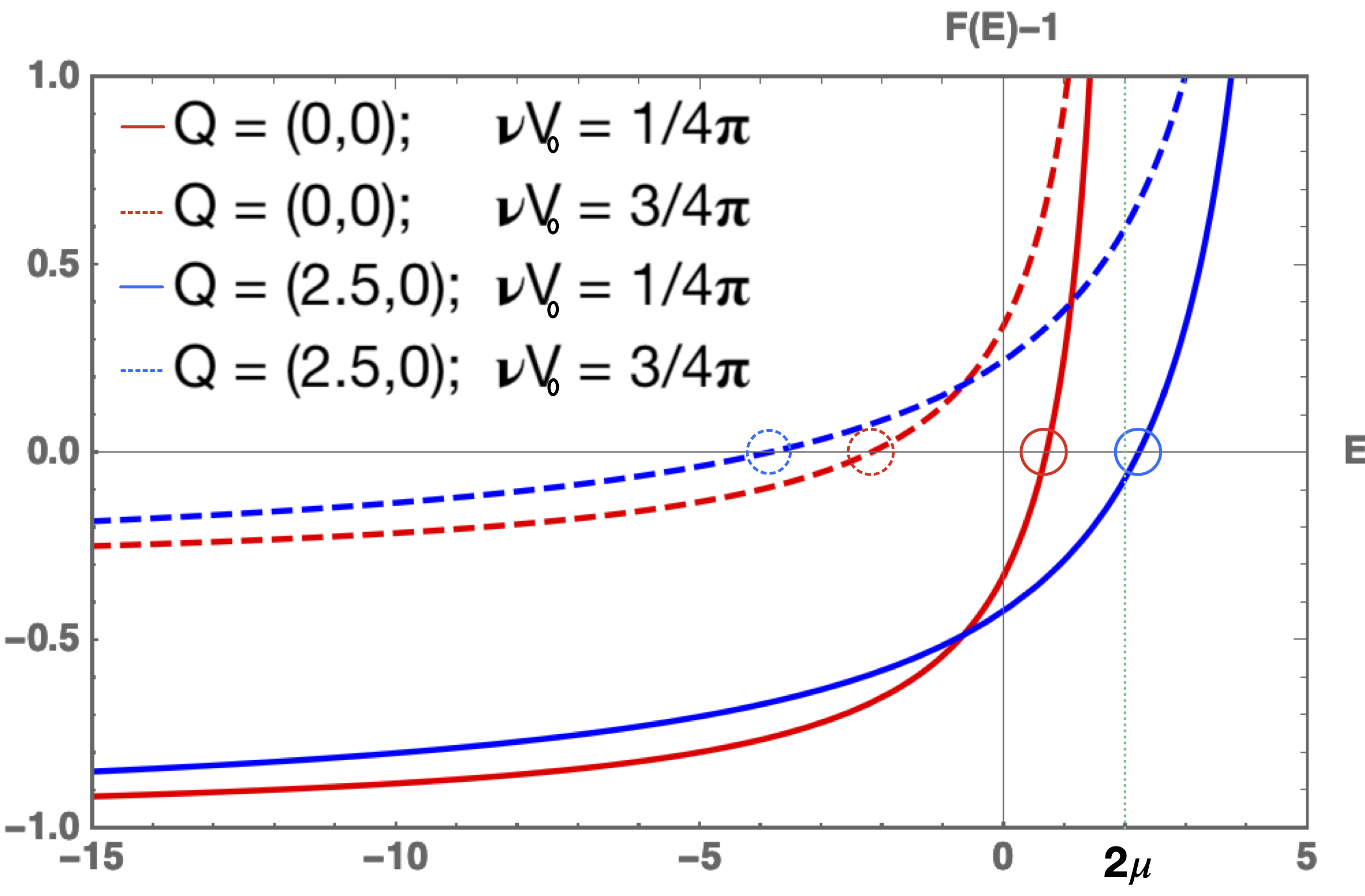} \qquad
    \caption{Plots of the function $F(E)-1$ with binding energy $E$ for two values of $\nu V_0 = 1/4 \pi, 3/4\pi$ and two values of $\bs Q = (0,0), (2.5,0)$. The dotted green line denotes $2 \mu$ and zeros of $F(E)-1$ (solid and dashed circles) denote two body bound state values. For small (large) $\nu V_0$, $\bs Q=0$ ($\bs Q \neq 0$) solution has a greater binding energy with respect to $2\mu$.   }
    \label{fig:FunctionOfV0}
\end{figure}

We now address the dependence of the finite $\bs Q$ transition on the interaction strength $V_0$. We show that for a given ratio $\Lambda/W$, finite momentum pairs are stabilized only above a critical value of $\nu V_{0c}$. We choose the continuum case for simplicity of illustration, but analogous arguments hold in the presence of a lattice. In the case of an isotropic $s$-wave pair potential, it is well known that even an infinitesimally small interaction ($\nu V_0$) can drive Cooper pairing with zero center-of-mass momentum. It is now established that a similar result holds for the uniform $\bs Q =0$ $d$-wave superconductor as well. This fact suggests that for weak enough interactions, the finite $\bs Q$ pairing gives way to homogeneous $\bs Q =0$ pairing for weak enough $\nu V_0$. To confirm this, we show in Fig.~\ref{fig:FunctionOfV0} a plot of $F(E) - 1$ with binding energy $E$ where $F(E)$ is defined in Eq.~\ref{Eq:ConditionForE}. We have chosen a Fermi energy with $\mu = 1$ and pairing within a narrow window $\mu \pm 0.2$.  A zero of the equation $F(E) - 1$ with $E<2\mu$ denotes pair binding. Here, an infinitesimally small $\nu V_0$, or any interaction value below a critical value $\nu V_{0c} \simeq 2/4\pi$ (solid curves in Fig.~\ref{fig:FunctionOfV0}), stabilizes a uniform $d$-wave superconductor (see also Methods). Finite momentum pairing in this limit acts to reduce the pair binding energy and is hence less favored. However, above the critical  $\nu V_{0c}$, finite momentum pairs (dashed curves in Fig.~\ref{fig:FunctionOfV0} with $\bs Q = (\pm 2.5, 0), (0, \pm 2.5)$)  gain more in pair binding energy than the uniform pairs. This leads to the stability of finite momentum pairing over uniform pairing.   \par 
\subsection{Variational wave function} 
\begin{figure}
    \centering
     \includegraphics[width=0.45\linewidth]{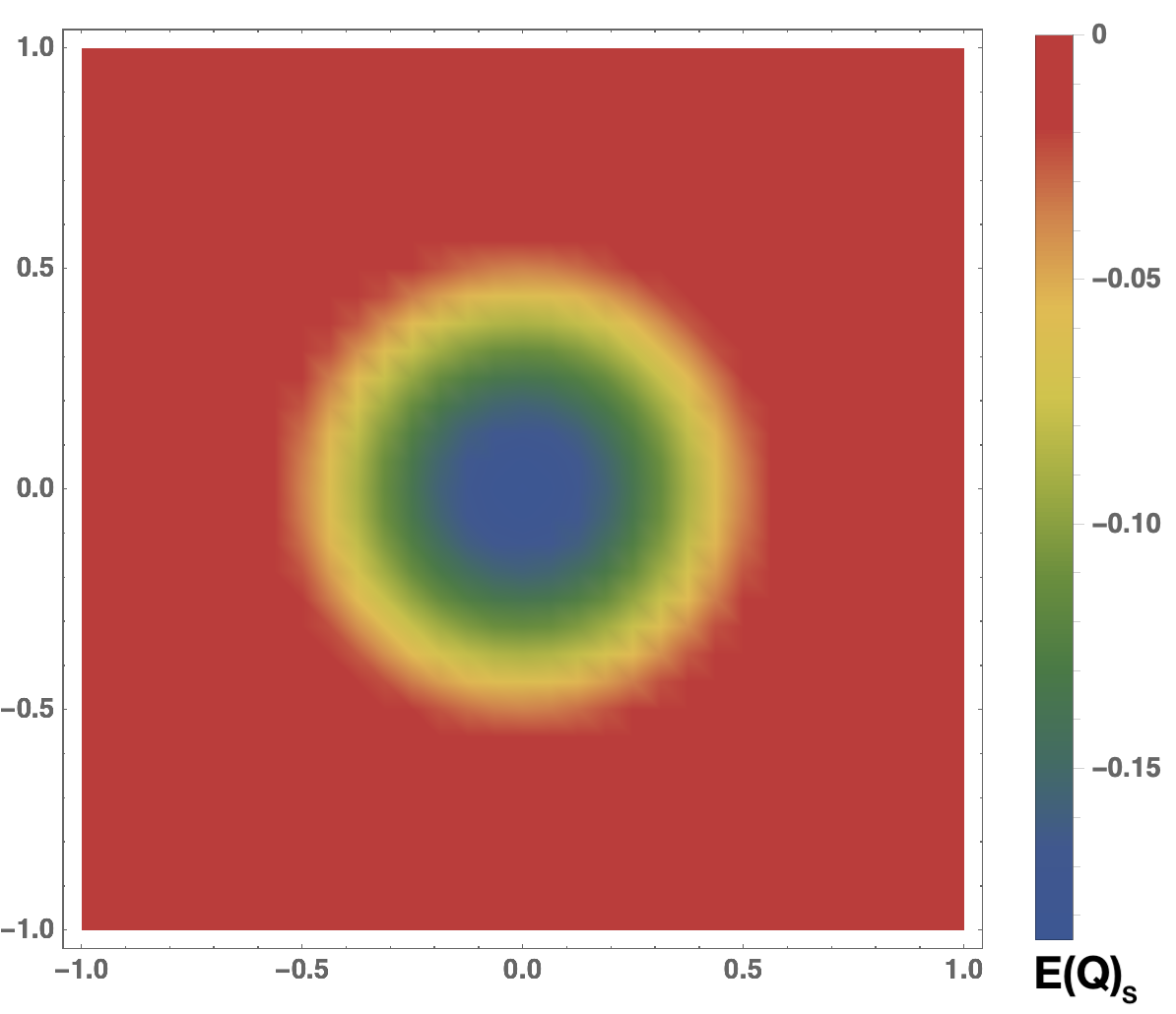} \qquad
     \includegraphics[width=0.45\linewidth]{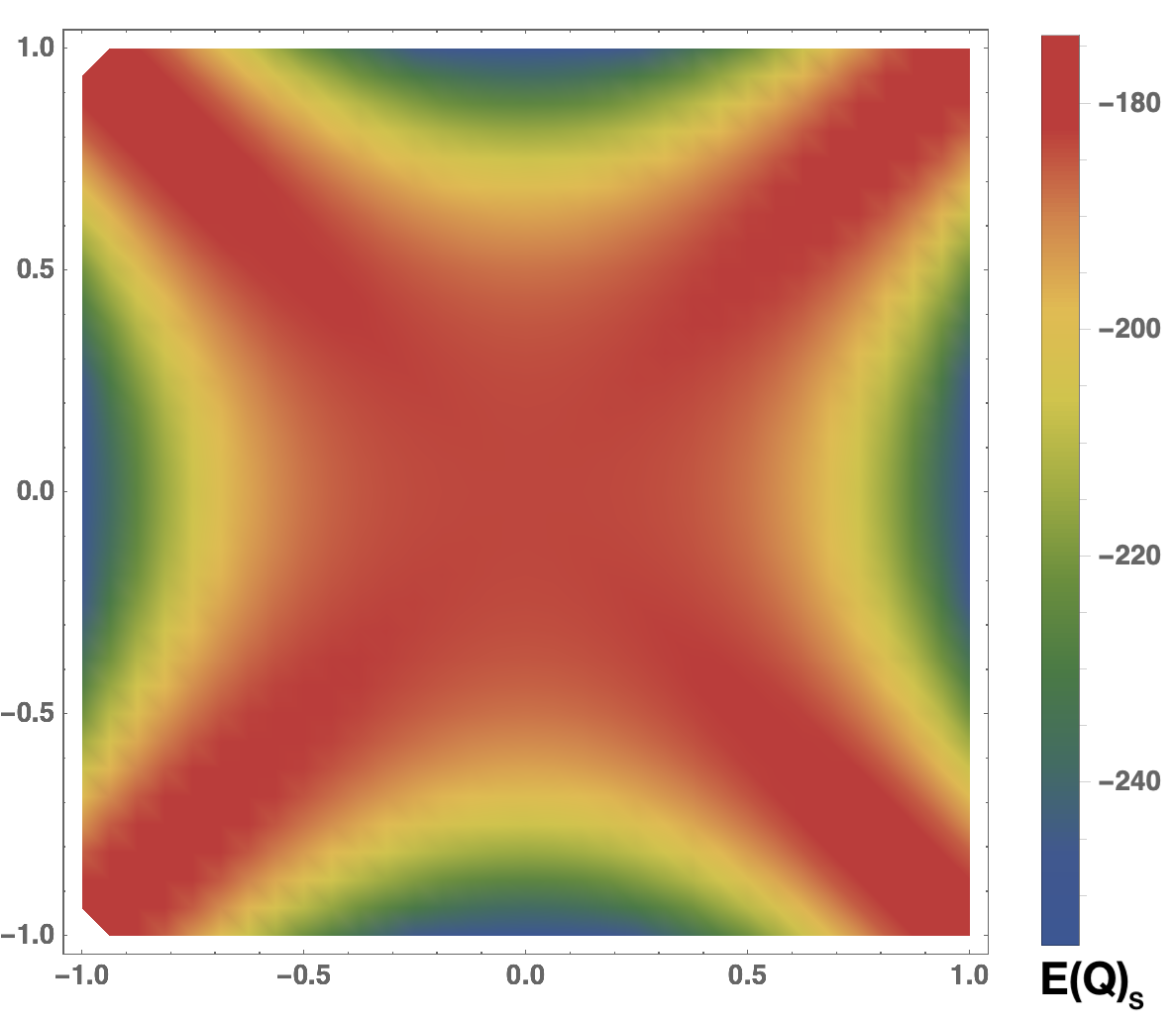} \qquad
      \includegraphics[width=0.45\linewidth]{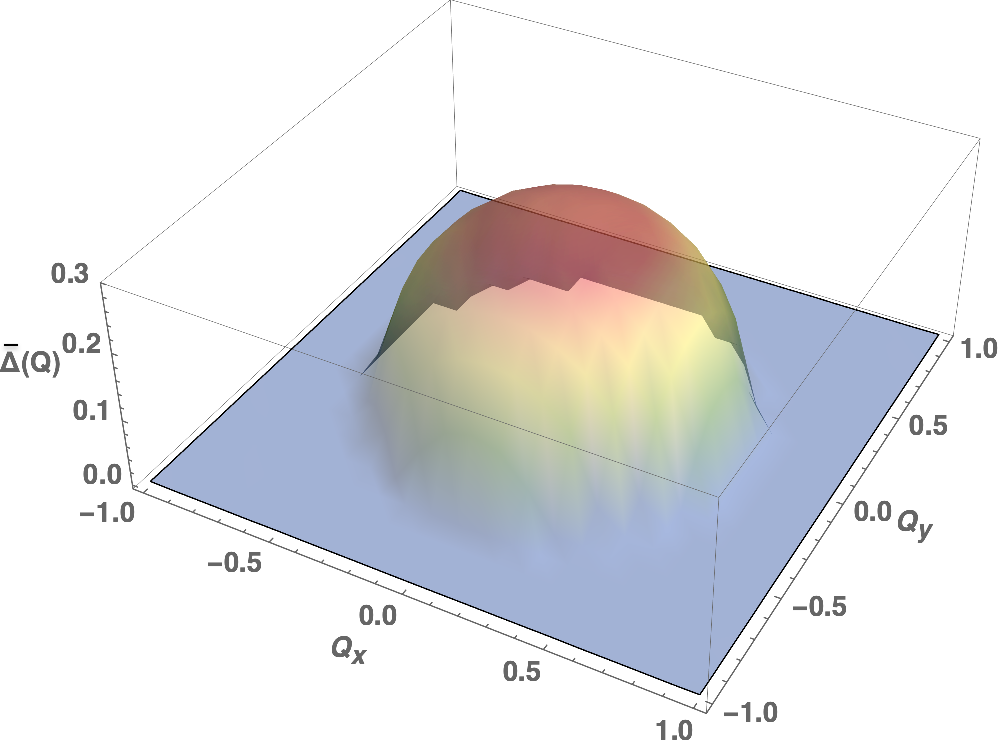} \qquad
     \includegraphics[width=0.45\linewidth]{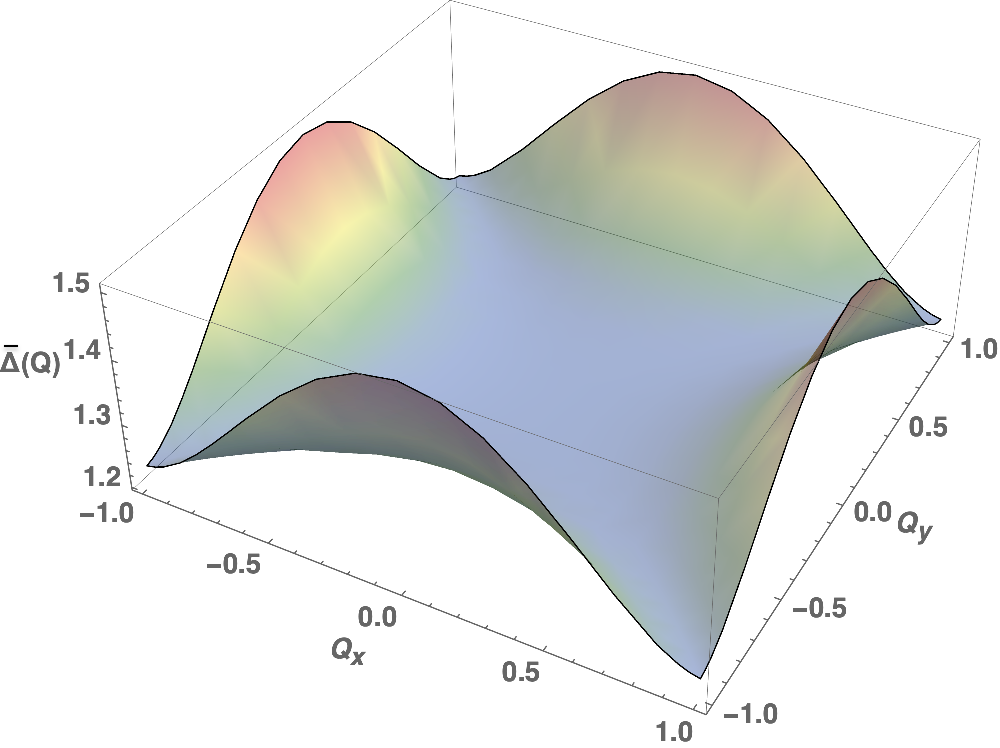} \qquad
    \caption{(Top row) Plots of the free energy $E(\bs Q)_S$ of the superconducting state as a function of center-of-mass momentum $\bs Q$ (units $\sqrt{\Lambda}$) for isotropic (left) and anisotropic (right) interactions. (Bottom row) Non-homogeneous order parameter $\bar{\Delta}(\bs Q)$ as a function of $\bs Q$ for isotropic (left) and anisotropic (right) interactions.  The chemical potential is set to $\mu=1$ and the interaction parameter is $\nu V_0 = 4/4\pi$.   }
    \label{fig:SCFreeEnergy-DwaveVsSwave}
\end{figure}

\begin{figure}
    \centering
     \includegraphics[width=0.48\linewidth]{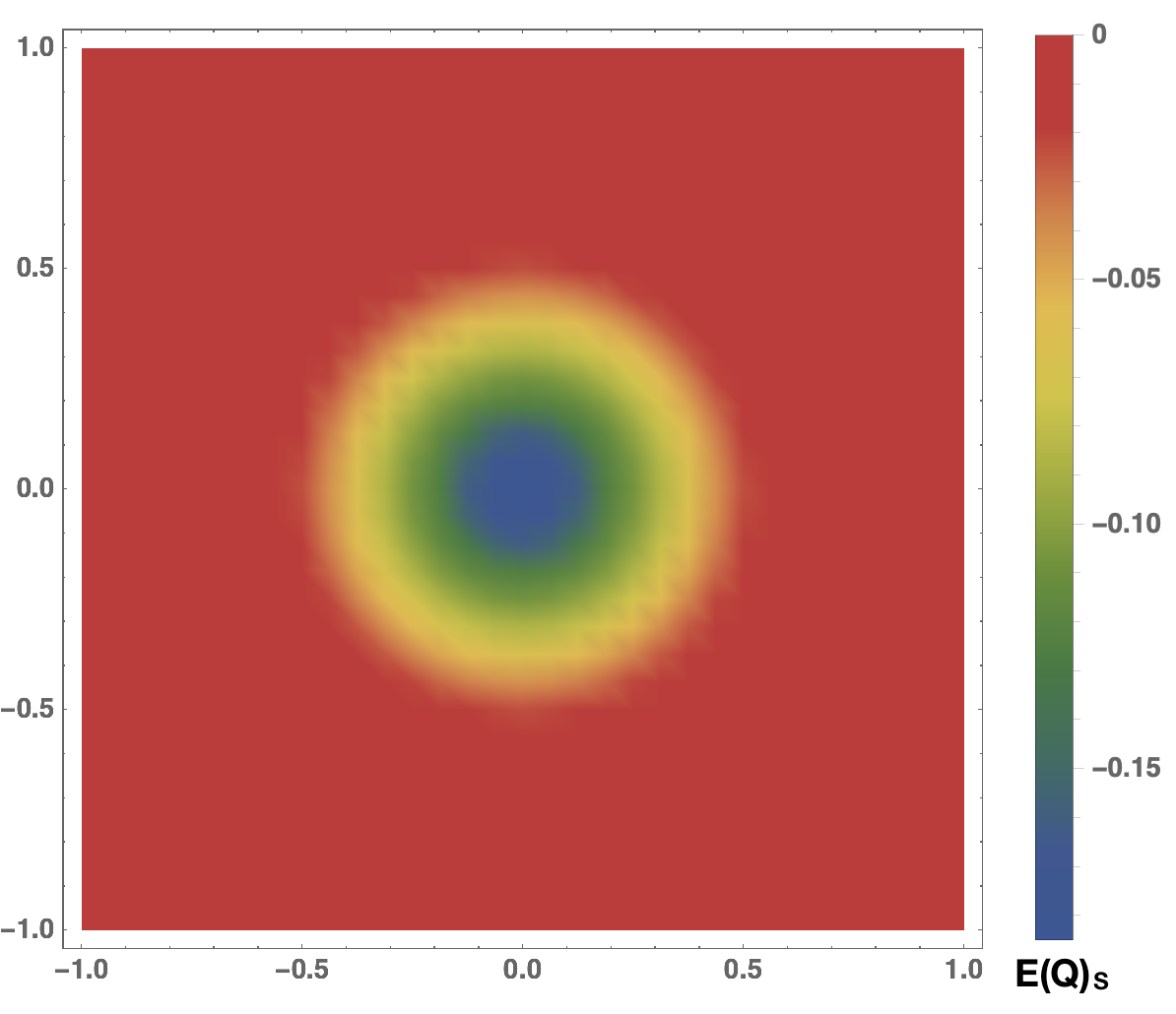} \qquad
     \includegraphics[width=0.48\linewidth]{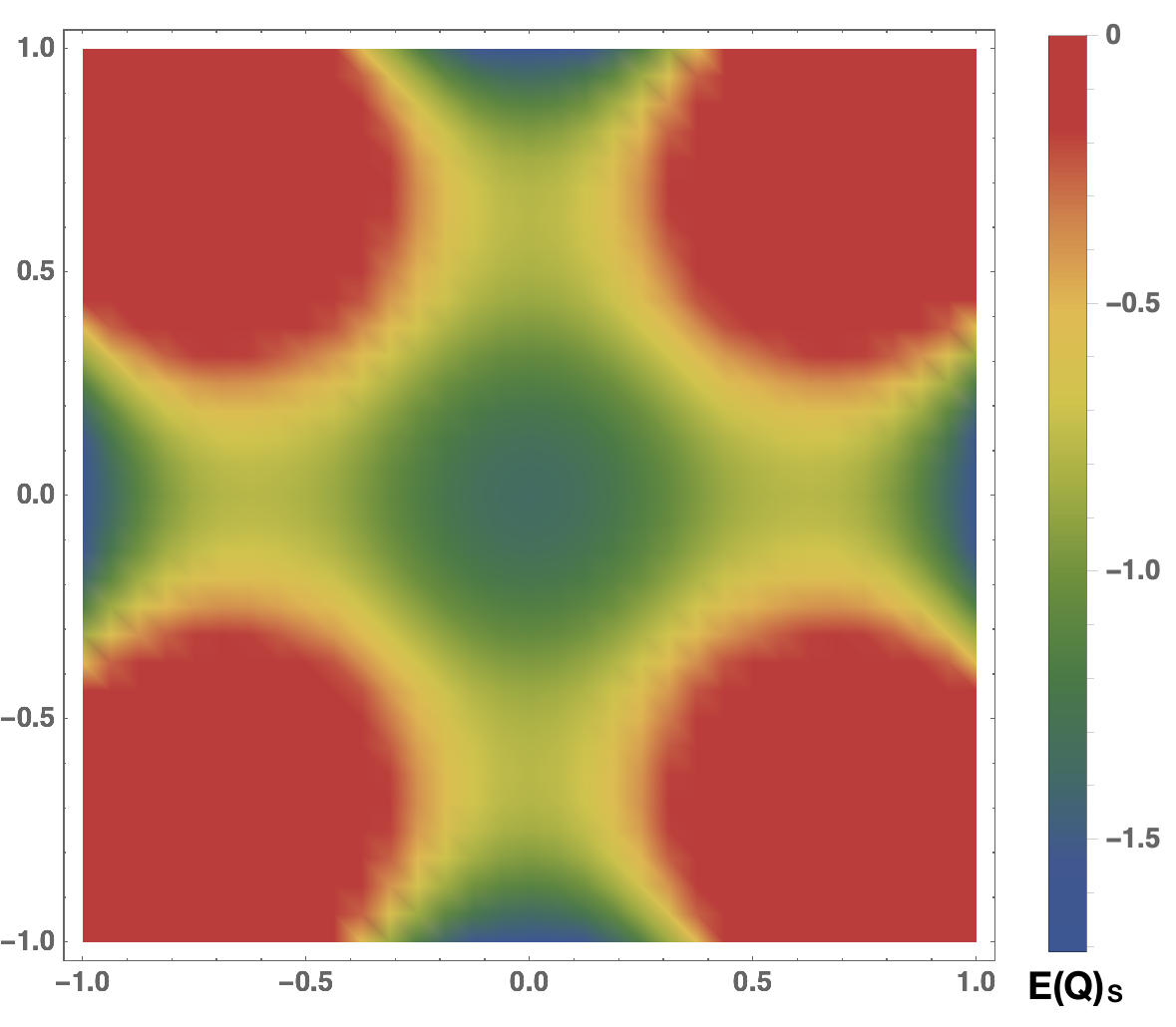} \qquad
     \includegraphics[width=0.48\linewidth]{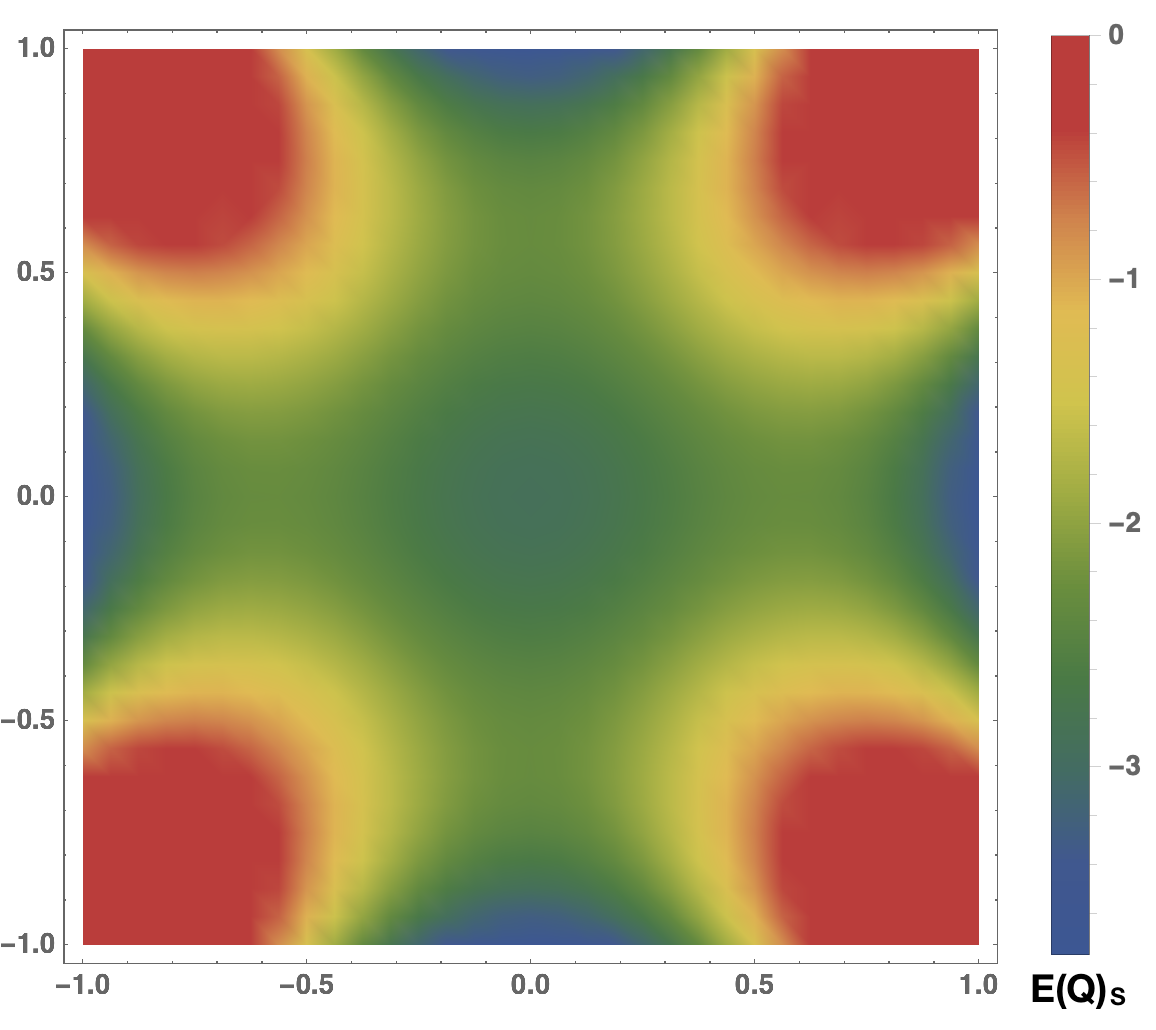} \qquad
    \caption{Plots of the free energy $E(\bs Q)_S$ of the superconducting state as a function of center-of-mass momentum $\bs Q$ (units $\sqrt{\Lambda}$) across the PDW transition. Top to Bottom panels: $\nu V_0 = 0.16/4 \pi, 0.2/ 4\pi, 0.24/ 4\pi$. The chemical potential is set to $\mu=1$.   }
    \label{fig:SCFreeEnergy-dwave-FunctionOfV0}
\end{figure}

The aforementioned conclusions of the two-body problem can be readily generalized to a many-body setting. To do this, we use the BCS variational wave function approach to evaluate and minimize the superconducting contribution to the free energy. In addition to the usual BCS-like variational parameters, we introduce the center-of-mass momentum coordinate $\bs Q$ as another variational parameter. A free energy minimum away from $\bs Q =0$ signals a transition to a many-body PDW phase.  We begin the analysis by writing the total Hamiltonian in second quantized notation
\beq
\mathscr H = \sum_{\bs k \sigma} \xi_{\bs k} c_{\bs k \sigma}^{\dagger}c_{\bs k \sigma} + \sum_{\bs k \bs k' \bs Q} V_{\bs k \bs k'}(\bs Q) c^{\dagger}_{\bs k + \bs Q/2~\uparrow } c^{\dagger}_{-\bs k + \bs Q/2~\downarrow} c_{-\bs k' + \bs Q/2~\downarrow} c_{\bs k' + \bs Q/2~ \uparrow}
\label{Eq:BCSHamiltonian}
\eeq
where $\xi_{\bs k}$ and $V_{\bs k \bs k'}(\bs Q)$ are the quasiparticle energy and interaction respectively, $c_{\bs k \sigma}^{\dagger}$ creates a quasiparticle with momentum $\bs k$ and spin $\sigma$,  $c_{\bs k + \bs Q/2~ \sigma}^{\dagger} c_{-\bs k + \bs Q/2~\bar{\sigma}}^{\dagger} $ creates a pair of spin-singlet quasiparticles with relative momentum $\bs k$ and center-of-mass momentum $\bs Q$ ($\bar{\sigma}$ is the spin-flip projection). We choose a variational ansatz for the PDW wave function with a single wave vector $\bs Q$  given by
\beq
|\Psi_{PDW} \rangle = \prod_{\bs k = \bs k_1..\bs k_M} \left( u_{\bs k} (
\bs Q) + v_{\bs k} (\bs Q) c_{\bs k + \bs Q/2~\uparrow}^{\dagger} c_{-\bs k + \bs Q/2~\downarrow}^{\dagger}  \right) |0\rangle. 
\label{Eq:Ansatz}
\eeq
Here $v(\bs Q)$ and $u(\bs Q)$ are variational parameters along with the center-of-mass momentum $\bs Q$ that must be determined by minimizing the free energy (see Methods), $|0\rangle$ is the vacuum state with no particles, and $\bs k_1.. \bs k_M$ are the momentum values in the band. The zero temperature free energy of the superconducting phase $E(\bs Q)_S = \langle \Psi_{PDW}| \mathscr{H} | \Psi_{PDW}\rangle $ 
is given by (see Methods)
\beq 
E(\bs Q)_S  = \frac{1}{2}\sum_{\bs k}  \bar{\xi}_{\bs k}(\bs Q) \left[ 1- \frac{\bar{\xi}_{\bs k}(\bs Q) }{E_{\bs k}(\bs Q)} \right] -  \frac{4 \bar{\Delta}(\bs Q)^2}{V_0}.
\label{Eq:Non-Homo-FE}
\eeq
where 
$\bar{\xi}_{\bs k} (\bs Q) \equiv \xi_{\bs k + \frac{\bs Q}{2}} +\xi_{-\bs k + \frac{\bs Q}{2}}$, $E_{\bs k}(\bs Q) \equiv \sqrt{\bar{\xi}_{\bs k} (\bs Q)^2 + 4 \Delta_{\bs k}(\bs Q)^2}$ and $\Delta_{\bs k}(\bs Q) = f_{\bs k}(\bs Q) \bar{\Delta}(\bs Q)$. Here $\bar{\Delta}(\bs Q)$ is a quantity independent of the relative momentum and is determined self-consistently using the formula
\beq
1 =V_0 \sum_{\bs k} \frac{f_{\bs k} (\bs Q)^2}{\sqrt{\bar{\xi}_{\bs k}(\bs Q)^2 + 4 f_{\bs k}(\bs Q)^2 \bar{\Delta}(\bs Q)^2} }.
\label{Eq:Non-Homo-Gap}
\eeq
The quantity $2 \Delta_{\bs k} (\bs Q)$ can be interpreted as the order parameter and takes the meaning of a superconducting gap while $E_{\bs k}(\bs Q)$ is the quasiparticle energy. Plots of the $\bs Q$ dependent  free energy and superconducting gaps ($\bar{\Delta}(\bs Q)$) in the continuum limit appear in Fig.~\ref{fig:SCFreeEnergy-DwaveVsSwave} for $n=1$ and $\mu=1$ with $\nu V_0 = 4/4\pi$. For isotropic interactions, the function $\bar{\Delta}(\bs Q)$ falls to zero for large $\bs Q$ uniformly in all directions; the free energy gain similarly approaches zero uniformly for large enough $\bs Q$ and the minimum occurs at $\bs Q =0$. This signals a homogeneous pairing state. However, for anisotropic interactions, the gap function increases along the anti-nodal directions ($(Q_x, Q_y) = (\pm 1, 0), (0, \pm 1)$).  Correspondingly, the free energy gain is maximized along these directions indicating instability of the zero $\bs Q$ phase towards a non-homogeneous solution. In the presence of a lattice, we can similarly show that the gap (free energy gain) is bounded with a maximum (minimum) occurring at $\bs Q \neq 0, \bs G$ when both $n=1,2$ harmonics are included, signalling a PDW state. These conclusions are similar to two-body results discussed in Fig.~\ref{fig:Compare}. In an analogous manner, Fig.~\ref{fig:SCFreeEnergy-dwave-FunctionOfV0} shows the evolution of the free energy as a function of $V_0$ for anisotropic interactions. For weak enough interaction strength, the free energy minimum occurs at $\bs Q =0$. The minimum then shifts away from the homogeneous solution upon tuning the interaction strength to larger values consistent with the two-body solution in Fig.~\ref{fig:FunctionOfV0}. 
\section{Conclusions} We presented an exact solution to a two-body Schrodinger wave equation that demonstrates the stability of a finite center-of-mass momentum Cooper pair over a homogeneous (zero momentum) $d$-wave pair. Currently, there is no known robust and generic mechanism, numerical or analytical, for why such modulated pairing is favored over a homogeneous superconductor in the presence of time reversal symmetry. This is despite widespread experimental evidence for fluctuating and static pair density waves in cuprates, Kagome metals and transition metal dichalcogenides.   The problem we defined here consists of two electrons subject to an anisotropic $d$-wave attractive potential dominant over a certain energy window.  Above a critical value of the interaction strength, non-zero center-of-mass momentum Cooper pairs save more binding energy in comparison with the homogeneous solution, hence stabilizing Cooper pairs of a PDW. The solution we provide in this paper is, therefore, an analog to the Cooper argument for a time reversal symmetry preserving fluctuating or static PDW. We have further demonstrated that these conclusions hold even in a many-body setting by explicitly evaluating the non-homogeneous superconducting gap and free energy using a BCS-like variational wave function ansatz for a PDW.  The stable finite momentum solution found here \textit{requires} an unconventional pairing interaction and cannot occur for isotropic $s$-wave interactions. In the latter case, Cooper pairs behave like point bosons, and must hence be uniform with a nodeless superconducting ground state.  Unconventional pairs, on the other hand, have non-trivial internal structure and do not behave like point bosons~\cite{deMelo2000}. This allows the ground state to stabilize finite momentum pairs with a nodal gap structure.  The aforementioned property provides a guiding principle to search for PDWs in superconductors with unconventional order parameters. Our result thus sets the stage for further analytical and numerical exploration of these intriguing phases of matter in correlated superconductors with unconventional pairing states.  \par
\section{Acknowledgements:} CS was primarily supported by the U.S. Department of Energy, Office of Science, Basic Energy Sciences, under Award No. DE-SC0018197 and additionally supported by the Robert A. Welch Foundation Grant No. C-1411. KY was supported by the National Science Foundation Grant No. DMR-1932796, and performed at the National High Magnetic Field Laboratory, which is supported by National Science Foundation Cooperative Agreement No. DMR-1644779, and the State of Florida. PWP thanks  DMR-2111379 for partial funding of this project. PJH is supported by the Department of Energy under Grant No. DE-FG02-05ER46236. LF acknowledges support by the European Union's Horizon 2020 research and innovation programme through the Marie Sklodowska-Curie grant SuperCoop (Grant No 838526). EWH was supported by the Gordon and Betty Moore Foundation EPiQS Initiative through the grants GBMF 4305 and GBMF 8691. 
 The authors thank Kavli Institute for Theoretical Physics and University of California, Santa Barbara for hospitality and the associated financial support from National Science Foundation under Grant No. NSF PHY-1748958. We benefited from discussions with Qimiao Si, and the organizers and attendees of the workshop ``Recent Progress in the Experimental and Theoretical Search for Pair-Density-Wave Order" where this work was conceived and initially presented.     \newline
\bibliography{PDW-Cooper}
\newpage
\section{Methods}
\textit{Deriving Eq.~\ref{Eq:ConditionForE}}: Substituting the spatial component of the wave function into the Schrodinger wave equation and writing in terms of Fourier components, we have
\beq
(\epsilon_{\bs k + \bs Q/2} + \epsilon_{-\bs k + \bs Q/2} - E)~g(\bs k) = - \sum_{\bs k'} V_{\bs k \bs k'} (\bs Q) g(\bs k')
\eeq
where $V_{\bs k, \bs k'}$ are the Fourier components of the two-body interaction which we assume to be separable of the form $V_{\bs k, 
\bs k'}(\bs Q) = f_{\bs k, \bs Q} f_{\bs k', \bs Q} $. For the purposes of our discussion, we choose $f_{\bs k, \bs Q} \equiv \frac{h_{\bs k - \bs Q/2  } + h_{\bs k + \bs Q/2  }}{2} $ where $h_{\bs k}$ is the $B_{1g}$ $d$-wave form factor.  \par
We can now solve for $g(\bs k)$ we have 
\beq
g(\bs k) = \frac{\sum_{\bs k'} V_{\bs k \bs k'}(\bs Q) g(\bs k')}{(-\epsilon_{\bs k + \bs Q/2} - \epsilon_{-\bs k + \bs Q/2} + E)}.
\label{Eq:gk}
\eeq
Multiplying both sides with $V_{\bs k, \bs k_1}(\bs Q)$ and summing to over $\bs k$ simplifies the expression. Noting that the interaction $V_{\bs k, \bs k'}(\bs Q)$ is assumed to be factorizable, momentum summations on the left and right hand sides cancel. Assuming a negative (attractive) potential strength $-|V_0|$, the Eq.~\ref{Eq:gk} above reduces to
\beq
1 = |V_0| \sum_{\bs k} \left[ \frac{f_{\bs k, \bs Q}^2}{- E + \epsilon_{\bs k + \bs Q/2} + \epsilon_{-\bs k + \bs Q/2}}\right].
\label{FinalEq}
\eeq
We can now solve Eq.~\ref{FinalEq} for the dispersion of binding energy $E(\bs Q)$. \\ \newline
\textit{Binding energy of homogeneous $d$-wave superconductor:} As we saw in the main text, for weak enough coupling $V_0$ in Eq.~\ref{Eq:ConditionForE}, the homogeneous $\bs Q =0$ solution in stable. Here we show that such a solution is exactly the conventional BCS $d$-wave superconductor. To prove this, we demonstrate that the $\bs Q =0$ solution to Eq.~\ref{Eq:ConditionForE} can occur at infinitesimally small coupling strength $V_0$. We begin by converting Eq.~\ref{Eq:ConditionForE} at $\bs Q =0$ into an energy integral that is non-zero only over a narrow energy window $\Omega$ around $\mu$:
\beq
\frac{1}{|\bar{V_0}|} = 2 \int_{\bar{\mu}}^{\bar{\mu} + \bar{\Omega}} \frac{(\bar{\xi} + \bar{\mu})^2 d\bar{\xi}}{-\bar E + 2 \bar{\xi}}
\eeq
Here the bar on top of each variable denotes normalization with respect to some high energy scale. The integral above can be performed to yield a non-linear equation for $\bar E$ given by
\beq
\frac{1}{|\bar{V_0}|} =\frac{1}{2} \left[ \bar{\Omega} (\bar{E} + 6 \bar{\mu} + \bar{\Omega}) + (\bar{E} + 2 \bar{\mu})^2 \tanh\left({\frac{\bar{\Omega}}{- \bar{E} + 2 \bar{\mu} + \bar{\Omega}}}\right) \right]
\eeq
This equation can be solved for $E<2\mu$ and a solution exists for infinitesimally small $|V_0|$. Hence we recover the binding energy of a conventional homogeneous $d$-wave superconductor.  
\\ \newline
\textit{Instability temperature $T_i$ of homogeneous $d$-wave superconductor:} Here we calculate the instability temperature $T_i$ for the homogeneous Cooper pair solution of Eq.~\ref{Eq:ConditionForE} and show that we can recover the weak coupling BCS-like formula. This can be done by taking the  $\bs Q\rightarrow 0$ limit of the finite temperature pair susceptibility using the same form factor appearing in Eq.~\ref{Eq:ConditionForE}. For a large and positive $\mu$ (large Fermi surface) the static zero momentum pair susceptibility is given by
\beq
\Pi(\bs q \rightarrow 0, \omega = 0) = \frac{1}{\beta \pi \Lambda^3} \sum_{\epsilon_n} \int_0^{\Lambda} \frac{4\pi k^5 dk}{(k^2 - \mu)^2 + \epsilon_n^2}
\eeq
where $\Lambda$ is an interaction cut-off, $
\beta$ the inverse temperature and $\epsilon_n$ is the Fermionic Matsubara frequency. Performing the Matsubara sum yeilds
\beq
\Pi(\bs q \rightarrow 0, \omega = 0) =\frac{1}{\beta \pi \Lambda^3} \int_0^{\Lambda} 4 \pi k^5 dk \frac{\tanh{\left(\frac{k^2 -\mu}{2 T} \right)}}{2 (k^2 - \mu) T}.
\eeq
The instability temperature $T_i$ (which equals the coherence temperature $T_c$ at weak coupling) can be solved by setting $\Pi(\bs q \rightarrow 0, \omega = 0)|_{T = T_i}- |V_0|^{-1} = 0$.  The non-linear equation can be solved for $T_i$ (for infinitesimal $|V_0|$)  and takes the approximate BCS-like form
\beq
T_i \simeq \frac{\Lambda}{4} \exp[\frac{-1}{\pi \nu |V_0|}]
\eeq
\\ \newline
\textit{Approximate $E(\bs Q)$ for $\bs Q$ along anti-nodal direction:} Here we derive an approximate formula for the binding energy $E(\bs Q)$ when $\bs Q$ is along the anti-nodal direction $\left[ \bs Q = (\pm q,0), (0, \pm q)\right]$. Substituting for $\bs Q = (q, 0)$ and neglecting the $q$ dependence of the denominators (this is a good approximation for large enough $q$), we get
\beq
F(E) \simeq -\frac{|V_0|}{2 \Lambda^3} \int_{\sqrt{2\mu}}^{\sqrt{2\mu + \Omega}} \frac{k (8 k^4 + q^4)}{E - 2 k^2+ 2\mu}dk.
\eeq
The integral above can be performed exactly and solved for $E(q)$ with $\Omega>0$. For small $\Omega$ and above the critical $|V_0|$ where zero momentum solution is unstable, the expression for $E(q)$ takes an illuminating form  
\beq
\bar E(q) \simeq 2 \bar{\mu} - \frac{1}{4} |\bar V_0| (\bar q^4 + 32 \bar{\mu}^2) \bar{\Omega}
\eeq
where all quantities are dimensionless and normalized by the appropriate factor of $\Lambda$. The expression above shows that a zero momentum solution gives rise to a finite momentum solution. The finite momentum is cut-off for large $q$ either by the energy cut-off or the lattice and the energy gain is thus bounded from below. \\ \newline
\textit{$E(\bs Q)$ for $\bs Q$ along the nodal points:} When $\bs Q$ is along the nodal direction, i.e, $\left[ \bs Q = (\pm q,\pm q), (\pm q, \mp q)\right]$, finite momentum pairs always reduce the gain pair binding energy; hence, they are unstable in comparison to the homogeneous $d$-wave superconductor. This can be seen readily by noting that the form factor in the interaction is independent of $\bs Q$ along the nodal regions; that is, $f_{\bs k, \bs Q} = f_{\bs k, 0}$ along the diagonals. The remaining  $\bs Q$ dependence of $F(E)$ occurs only in its denominator like in the $s$-wave case.  Hence, finite momentum pairs always cost energy along the diagonal direction.    \\ \newline
\textit{Many body free energy and non-homogeneous gap using variational ansatz:} Here we provide details leading to the expressions for the variational free energy and non-homogeneous gap functions Eqs.~ \ref{Eq:Non-Homo-FE}, \ref{Eq:Non-Homo-Gap}. To derive these equations, we take the expectation value of the total Hamiltonian Eq.~\ref{Eq:BCSHamiltonian} with respect to the ground state ansatz $|\Psi_{PDW}\rangle$ Eq.~\ref{Eq:Ansatz}. The kinetic energy term yields
\beq
\langle \Psi_{PDW} | \sum_{\bs k \sigma} \xi_{\bs k} c_{\bs k \sigma}^{\dagger} c_{\bs k \sigma} | \Psi_{PDW} \rangle = \sum_{\bs k } \left( \xi_{\bs k + \bs Q/2} + \xi_{-\bs k + \bs Q/2} \right) v_{\bs k}(\bs Q)^2
\label{Eq:KEPart}
\eeq
where the two terms on the right hand side with relative momenta $\bs Q$ correspond to the two spin projections ($\uparrow, \downarrow$) respectively. Similarly the interaction term yields
\beq\nonumber
\langle \Psi_{PDW} | \sum_{\bs k \bs k' \bs Q} V_{\bs k \bs k'}(\bs Q) c^{\dagger}_{\bs k + \bs Q/2~\uparrow } c^{\dagger}_{-\bs k + \bs Q/2~\downarrow} c_{-\bs k' + \bs Q/2~\downarrow} c_{\bs k' + \bs Q/2~ \uparrow} | \Psi_{PDW} \rangle &=& \\ 
&&\!\!\!\!\!\!\!\!\!\!\!\!\!\!\!\!\!\!\!\!\!\!\!\!\!\!\!\!\!\!\!\!\!\!\!\!\!\!\!\!\!\!\!\!\!\!\!\!\!\!\!\!\!\!\!\!\!\!\!\!\!\!\!\!\!\!\!\!\!\!\!\!\!\!\!\!\! \sum_{\bs k, \bs k'} V_{\bs k \bs k'}(\bs Q) u_{\bs k}(\bs Q) v_{\bs k} (\bs Q) u_{\bs k'}(\bs Q) v_{\bs k'}(\bs Q). 
\label{Eq:InteractionPart}
\eeq
The wave function components $u_{\bs k}(\bs Q),v_{\bs k}(\bs Q) $ satisfy the constraint $u_{\bs k}(\bs Q)^2 + v_{\bs k}(\bs Q)^2 =1$ and can be parameterized as $u_{\bs k} (\bs Q) = \sin \theta_{\bs k, \bs Q} $ and $v_{\bs k}(\bs Q)= \cos \theta_{\bs k, \bs Q} $. We now minimize the total expectation value $E(\bs Q)_S = \langle \Psi_{PDW}| \mathscr{H} | \Psi_{PDW}\rangle $ with respect to the parameter $\theta_{\bs k, \bs Q}$ and we obtain the condition
\beq
\tan2\theta_{\bs k, \bs Q} = \frac{-2 \Delta_{\bs k}(\bs Q)}{\left( \xi_{\bs k + \bs Q/2} + \xi_{-\bs k + \bs Q/2} \right)} \equiv \frac{-2 \Delta_{\bs k}(\bs Q)}{ \bar{\xi}_{\bs k}(\bs Q)},
\eeq
so that $ 2 u_{\bs k}(\bs Q) v_{\bs k}(\bs Q) = \sin2\theta_{\bs k, \bs Q} = \frac{2 \Delta_{\bs k} (\bs Q)}{E_{\bs k}(\bs Q)}$ and $ v_{\bs k}(\bs Q)^2 - u_{\bs k}(\bs Q)^2 = \cos2\theta_{\bs k, \bs Q} = -\frac{\bar{\xi}_{\bs k} (\bs Q)}{E_{\bs k}(\bs Q)}$.  As stated in the main text, here we have defined the quantities 
\beq
\Delta_{\bs k}(\bs Q) &\equiv& -\frac{1}{2}\sum_{\bs k'} V_{\bs k \bs k'}(\bs Q) \sin 2\theta_{\bs k, \bs Q} = - \sum_{\bs k'} V_{\bs k\bs k'}(\bs Q) u_{\bs k' }(\bs Q) v_{\bs k'} (\bs Q)\\   
E_{\bs k}(\bs Q) &\equiv&  \sqrt{\bar{\xi}_{\bs k}(\bs Q)^2 + 4 \Delta_{\bs k}(\bs Q)^2}.
\eeq
Using these definitions and the normalization relation $u_{\bs k}(\bs Q)^2 + v_{\bs k}(\bs Q)^2 =1$, we can obtain explicit expressions for the variational parameters $u_{\bs k}(\bs Q), v_{\bs k}(\bs Q)$ as 
\beq \nonumber
v_{\bs k}(\bs Q) &=& \sqrt{\frac{1}{2}\left(1 - \frac{\bar{\xi}_{\bs k}(\bs Q)}{E_{\bs k}(\bs Q)}\right)}\\
u_{\bs k}(\bs Q) &=& \sqrt{\frac{1}{2}\left(1 + \frac{\bar{\xi}_{\bs k}(\bs Q)}{E_{\bs k}(\bs Q)}\right)}.
\label{Eq:VariationalParameters}
\eeq
To arrive at the condition for the non-uniform gap function (Eq.~\ref{Eq:Non-Homo-Gap} main text), we substitute the variational parameters into the definition for $\Delta_{\bs k}(\bs Q)$ above. After further utilizing the form of the interaction in Eq.~\ref{Eq:Cases} we obtain the gap equation
\beq
\Delta_{\bs k} (\bs Q) = \frac{V_0}{2} \sum_{\bs k'} \frac{2 \Delta_{\bs k'}(\bs Q)}{E_{\bs k'}(\bs Q)} f_{\bs k}(\bs Q) f_{\bs k'}(\bs Q).
\eeq
The ansatz that solves the above equation takes the form $\Delta_{\bs k}(\bs Q) = f_{\bs k}(\bs Q) \bar{\Delta}(\bs Q)$ where $\bar{\Delta}(\bs Q)$ is independent of the relative momentum $\bs k$. Substituting the ansatz into $\Delta_{\bs k} (\bs Q)$, the gap equation simplifies to Eq.~\ref{Eq:Non-Homo-Gap}
\beq
1 =V_0 \sum_{\bs k} \frac{f_{\bs k} (\bs Q)^2}{\sqrt{\bar{\xi}_{\bs k}(\bs Q)^2 + 4 f_{\bs k}(\bs Q)^2 \bar{\Delta}(\bs Q)^2} }.
\eeq
We can further easily obtain the superconducting contribution to the free energy by 
utilizing Eqs.~\ref{Eq:Non-Homo-Gap}, \ref{Eq:KEPart}, \ref{Eq:InteractionPart} and \ref{Eq:VariationalParameters}. Substituting Eq.~\ref{Eq:VariationalParameters} into Eqs.~\ref{Eq:KEPart},~\ref{Eq:InteractionPart} and utilizing Eq.~\ref{Eq:Non-Homo-Gap}, we obtain
\beq 
E(\bs Q)_S =\langle \Psi_{PDW}| \mathscr{H} | \Psi_{PDW}\rangle  = \frac{1}{2}\sum_{\bs k}  \bar{\xi}_{\bs k}(\bs Q) \left[ 1- \frac{\bar{\xi}_{\bs k}(\bs Q) }{E_{\bs k}(\bs Q)} \right] -  \frac{4 \bar{\Delta}(\bs Q)^2}{V_0},
\eeq
which is Eq.~\ref{Eq:Non-Homo-FE} of the main text. 
\newpage

\end{document}